\documentclass[proof]{pasj01}

\Received{$\langle$received$\rangle$}
\Accepted{$\langle$accepted$\rangle$}
\Published{$\langle$publication date$\rangle$}

\begin{document}

\title{Suzaku observations of a shock front tracing the western edge of the giant radio halo in the Coma Cluster}

\author{Yuusuke \textsc{Uchida}\altaffilmark{1,2}}
\altaffiltext{1}{Institute of Space and Astronautical Science (ISAS), JAXA, 3-1-1 Yoshinodai, Chuo-ku, Sagamihara, Kanagawa, 252-5210 Japan}
\altaffiltext{2}{Department of Physics, Graduate School of Science, University of Tokyo, 7-3-1 Hongo, Bunkyo, Tokyo 113-0033, Japan}
\email{uchida@astro.isas.jaxa.jp}

\author{Aurora \textsc{Simionescu}\altaffilmark{1}}
\author{Tadayuki \textsc{Takahashi}\altaffilmark{1,2}}
\author{Norbert \textsc{Werner}\altaffilmark{3,4}}

\altaffiltext{3}{KIPAC, Stanford University, 452 Lomita Mall, Stanford, CA 94305, USA}
\altaffiltext{4}{Department of Physics, Stanford University, 382 Via Pueblo Mall, Stanford, CA  94305-4060, USA}

\author{Yuto \textsc{Ichinohe}\altaffilmark{1,2}}
\author{Steven W. \textsc{Allen}\altaffilmark{3,4,5}}
\author{Ondrej \textsc{Urban}\altaffilmark{3,4,5}}

\altaffiltext{5}{SLAC National Accelerator Laboratory, 2575 Sand Hill Road, Menlo Park, CA 94025, USA}

\author{Kyoko \textsc{Matsushita}\altaffilmark{6}}
\altaffiltext{6}{Department of Physics, Tokyo University of Science, 1-3 Kagurazaka, Shinjyuku-ku, Tokyo 162-8601, Japan}


\maketitle

\begin{abstract}

We present the results of new Suzaku observations of the Coma Cluster, the X-ray brightest, nearby, merging system hosting a well studied, typical giant radio halo. 
It has been previously shown that, on the western side of the cluster, the radio brightness shows a much steeper gradient compared to other azimuths. XMM-Newton and Planck revealed a shock front along the southern half of the region associated with this steep radio gradient, suggesting that the radio emission is enhanced by particle acceleration associated with the shock passage. Suzaku demonstrates for the first time that this shock front extends northwards, tracing the entire length of the western edge of the Coma radio halo. The shock is detected both in the temperature and X-ray surface brightness distributions and has a Mach number of around $\mathcal{M}\sim1.5$. The locations of the surface brightness edges align well with the edge of the radio emission, while the obtained temperature profiles seem to suggest shocks located 125--185 kpc further out in radius. In addition, the shock strengths derived from the temperature and density jumps are in agreement when using extraction regions parallel to the radio halo edge, but become inconsistent with each other when derived from radial profiles centred on the Coma Cluster core. It is likely that, beyond mere projection effects, the geometry of the shock is more complex than a front with a single, uniform Mach number and an approximately spherically symmetric shape.

\end{abstract}

\section{Introduction}

Clusters of galaxies grow through mergers with other (sub-)clusters and by accreting material from the surrounding large-scale structure filaments. 
Major mergers between systems of comparable masses release tremendous amounts of energy into the intracluster medium (ICM). These events drive shocks and turbulence in the ICM, thereby heating the thermal diffuse gas, accelerating particles to relativistic energies, and amplifying the intra-cluster magnetic fields (see reviews by e.g. \cite{b7,b6} and references therein).

Evidence for these mergers can be readily identified at a variety of wavelengths. In the hot, diffuse ICM, merger shocks create discontinuities in the temperature, density and pressure profiles that can be measured through X-ray observations. More recently, shocks have also been detected using the Sunyaev Zel'dovich (SZ) effect, whose strength is directly proportional to the thermal pressure of the X-ray emitting plasma (e.g. \cite{b1}).

In addition, particles accelerated to relativistic energies during the mergers emit synchrotron radiation in the radio band. Depending mainly on their morphology and location with respect to the X-ray peak, radio features in galaxy clusters are categorised either as radio haloes or relics (\cite{b6,b17}). Radio haloes are relatively symmetric, usually unpolarised, and cover the central parts of galaxy clusters, on scales of typically 1 - 2 Mpc. There are two main hypotheses to explain their origin: (re)acceleration of electrons up to relativistic energies due to the turbulence induced by cluster mergers (e.g. \cite{b35}), or collisions between thermal ions and relativistic protons in the ICM (\cite{b36}). Radio relics are typically elongated in the tangential direction and located near the clusters' periphery. Their origin is thought to be connected to particle acceleration at shock fronts driven by the merger (e.g. \cite{b5}, \cite{ogr13}).

Multi-wavelength studies of the thermal and non-thermal contents of clusters of galaxies using X-ray, SZ, and radio data can therefore reveal important information regarding the mechanisms through which particles are accelerated (and the ICM is heated) during major cluster mergers.

\begin{table*}
 \centering
 \begin{minipage}{100mm}
\caption{{\it Suzaku} observation Logs}
\begin{tabular}{cccccc} \hline
&Obs. ID & Obs. Date & R.A. & Dec. & Exposure \\
&&&&&(ks) \\ \hline
W2 & ae808090010 & 2013-07-02 & 12 56 43.97 & +28 16 25.7 & 12 \\
W3 & ae808091010 & 2013-07-02 & 12 56 15.34 & +28 20 14.6 & 15 \\ \hline
W1 & ae802084010 & 2007-06-21 & 12 57 22.27 & +28 08 25.1 & 29 \\
W4 & ae806025010 & 2011-12-09 & 12 55 53.76 & +28 30 38.9 & 17 \\ \hline
\end{tabular}
\label{tab:obssum}
\end{minipage}
\end{table*}


As one of the nearest, brightest, most massive merging systems, the Coma Cluster has been the target of a wealth of observations covering a wide range of the electromagnetic spectrum. The cluster is known to host both a giant radio halo (Coma C) and a radio relic towards the south-west (\cite{b29,bal81}). X-ray observations have revealed the presence of a shock front associated with the radio relic (\cite{b45,ogr13b}). The giant radio halo exhibits a relatively sharp edge observed at 352 MHz with the Westerbork Synthesis Radio Telescope (WSRT, \cite{b11}).

Using SZ observations of the Coma Cluster, \cite{b1} indicated the existence of pressure jumps toward the south-west and south-east of the Coma Cluster centre, whose positions coincide with the edge of the WSRT giant radio halo. These pressure jumps are also associated with jumps in the X-ray temperature, especially along the western azimuth, and are therefore interpreted as weak shocks. The Mach number calculated from the pressure discontinuity along the western direction reported by \cite{b1} is $\mathcal{M}_\mathrm{west} = 2.03^{+0.09}_{-0.04}$. Using {\it Suzaku} observations, \cite{b2} also report the detection of a temperature jump likely associated with a weak shock along the western side of the Coma Cluster, however located at a different radius and azimuth than the shock reported by \cite{b1}.

In this paper, we present the results from two new {\it Suzaku} observations located along the western edge of the Coma radio halo, which were obtained in addition to the data presented in \cite{b2}. We have assumed a $\Lambda$ cold dark matter cosmology with $\Omega_\mathrm{m} = 0.27, \Omega_\Lambda = 0.73$ and $H_0 = 70 \mathrm{\ km\ s}^{-1}\mathrm{Mpc}^{-1}$. The Coma Cluster redshift is $z = 0.0231$ (\cite{b3}), and one arcmin corresponds to 28 kpc. We define the virial radius as $r_{200}$, the radius within which the mean enclosed density is 200 times the critical density of the Universe at the redshift of the cluster. For the Coma Cluster, $r_{200}$ corresponds to 2 Mpc (70 arcmin), with consistent values derived both from the $Y_{X} = M_{gas} \times kT$ parameter (\cite{b1}) and from weak lensing analysis (\cite{b4}).


\section[]{Observations and Data Reduction}

We analyzed four Suzaku pointings located towards the west (W) from the Coma Cluster centre, with a total exposure time of 73~ks. We denote these data sets as W1, W2, W3 and W4, with increasing index indicating a larger distance of the aim-point with respect to the centre of the Coma Cluster. W1 and W4 were part of the mosaic analysed by \cite{b2}, while W2 and W3 are new observations performed by {\it Suzaku} in 2013 July (AO-8). The observation logs are summarized in Table \ref{tab:obssum}.

We reduced the data from the X-ray Imaging Spectrometers (XIS) 0, 1 and 3 using HEAsoft tools (version 6.13) and the calibration database (CALDB) 2015-01-05.   
In addition to the standard screening using the criteria recommended by the XIS team, we selected only observing periods with the geomagnetic cut-off rigidity COR $>$ 6 GV.
For the XIS1 data taken after 2011 June 1st, when the charge injection level was increased, we excluded two adjacent columns on each side of the charge injected columns. 
We also excluded the columns in Segment A of XIS0 that were affected by an anomaly in 2009 June.

Redistribution Matrix Files (RMF) and Ancillary Response Files (ARF) were creating using the FTOOLS tasks {\tt xisrmfgen} and {\tt xissimarfgen}.

\begin{figure*}
\begin{center}
\includegraphics[scale=0.5]{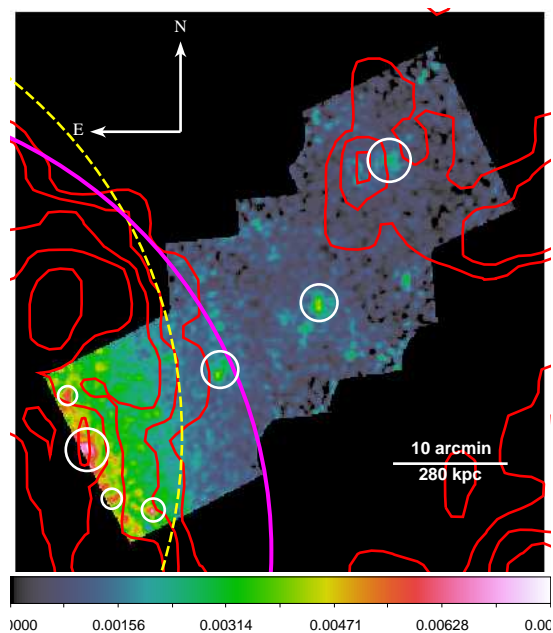}
\end{center}
\caption{Surface brightness map of the four Suzaku pointings analysed in this work, using the 0.7-7.0 keV energy band. The image has been corrected for instrumental background and vignetting. White circles indicate the point sources excluded from the analysis. 352 MHz WSRT radio contours from \cite{b11} are shown in red. The contours start at 4mJy/beam and increase in intervals of 3 mJy/beam. The magenta circle marks a distance of 40 arcmin from the Coma Cluster centre, while the yellow circle denotes a distance of 40 arcmin from the "analysis centre" chosen to match the radio halo edge.}
\label{fig:w1234}
\end{figure*}

\section[]{Background Analysis}

Precise background subtraction is important for the spectral analysis of cluster outskirts, where the surface brightness is low.
We, thus, carefully estimated both the non-X-ray background and cosmic X-ray backgrounds as described below.

\subsection{Non-X-ray Background}

The non-X-ray background (NXB) spectrum corresponding to each detector and each spectral extraction region was created from night Earth observations using the task {\tt xisnxbgen}. \cite{b13} demonstrated that {\tt xisnxbgen} reproduces the NXB spectrum within a 5 \% accuracy.

In order to further check the reliability of the NXB spectra for the newest observations (w2 and w3), we employed the method suggested by \cite{b12}. Briefly, for the front-illuminated (FI) devices (XIS 0 and 3), we measured the strength of the Ni {\small I} K $\alpha$ instrumental line at 7.47 keV in the NXB and the observed (non-background corrected) spectra by fitting these with a power-law plus Gaussian model in a narrow energy band centred around this emission line. 
We find that the normalisations of the Gaussian model agree between the observed and the corresponding NXB spectra, with ten out of twelve data sets (for the extraction annuli described in Section \ref{specan}) showing an agreement within the $1\sigma$ confidence interval, and two data sets showing an agreement between 1--2$\sigma$, in accordance with the expected statistical dispersion.
For the back-illuminated device (XIS1), which does not show instrumental emission lines at high energies, we compared the count rate between the observed and NXB spectra in the 8-10 keV range and also find a good agreement given the expected statistical dispersion.
We concluded that the NXB files created by {\tt xisnxbgen} are a good description of the instrumental background corresponding to our observations.

\subsection{The X-ray backgrounds}\label{sect:xrb}

The X-ray background components are the Cosmic X-ray Background (CXB) due to the sum of distant, unresolved point sources, the Galactic Halo emission (GH) and the Local Hot Bubble (LHB). We adopt the same background model parameters determined by \cite{b2}, using regions in the Suzaku mosaic observations of the Coma Cluster that were located just beyond its virial radius. 
The effect of Galactic absorption was accounted for using the {\it phabs} model with the column density $N_\mathrm{H}$ fixed at the average value computed from the Leiden-Argentine-Bonn radio HI survey (\cite{b21}) for the coordinates corresponding to each {\it Suzaku} field of view.

The CXB was modelled as a single power-law modified by the Galactic absorption. The power law index and normalization reported by \cite{b2} are $\Gamma_\mathrm{CXB} = 1.50$ and $\mathit{Norm}_\mathrm{CXB} = 1.15 \times 10^{-3}$ $\mathrm{keV}^{-1} \mathrm{cm}^{-2} \mathrm{s}^{-1}$ per $20^2 \pi$ arcmin$^2$ at 1 keV, respectively.
The GH was modelled as an absorbed thermal plasma with temperature $kT_\mathrm{GH} = 0.25$ keV, normalisation $\mathit{Norm}_\mathrm{GH} = 2.9 \times 10^{-3}$, and 0.3 Solar metallicity, while the LHB model consisted of an unabsorbed thermal plasma with $kT_\mathrm{LHB} = 0.10$ keV, $\mathit{Norm}_\mathrm{LHB} = 1.6 \times 10^{-3}$, and Solar metallicity assuming the abundance table of \cite{b24}. The normalizations of the thermal components correspond to $\int n_\mathrm{e}n_\mathrm{H} \mathrm{d} V \times( 10^{-14}/(4\pi[D_A(1+z)]^2)) \mathrm{\ cm}^{-5}$ per $20^2 \pi$ arcmin$^2$ , where $D_A$ is angular diameter distance, $n_\mathrm{e}$ and $n_\mathrm{H}$ are electron and hydrogen densities, respectively, and $z$ is the redshift.

\section{Analysis and Results}

To investigate the shock along the western azimuth, we carried out temperature and surface brightness analyses, considering the energy range between $0.7 - 7$ keV.

\subsection{X-ray images}

To obtain a mosaic image of the region of interest, we combined the images from all XIS detectors and all pointings listed in Table \ref{tab:obssum}. We excluded anomalously bright pixels as well as the regions located within 30 arcsec from the detector edges. Images of the instrumental background (NXB) scaled to the corresponding exposure time for each detector and each observation were subtracted from the raw count maps. To take into account the vignetting effect, we created flat-field maps using the task {\tt xissim}, which performs a ray-tracing Monte Carlo simulation of the X-ray telescope (XRT) and XIS\footnote{see https://heasarc.gsfc.nasa.gov/docs/suzaku/analysis/expomap.html}. The input photon list for {\tt xissim} was that representing a spatially uniform emission with a spectrum corresponding to the best-fit average model of the cluster emission and sky background obtained from the central $6.5^\prime$ of each respective pointing. Using this model spectrum, the ray tracing then accounts appropriately for the energy dependence of the vignetting effect.
The vignetting corrected image was created by dividing the background subtracted XIS image by the flat-field map.
The obtained image is shown in Figure \ref{fig:w1234}.
In the same figure, we indicate the 352 MHz WSRT radio contours (\cite{b11}) in red.

For the temperature and surface brightness profiles presented in the following sections, we have considered annuli with two different centres: one is the cluster centre, which we adopt as $(\alpha, \delta) = (12\mathrm{h}59\mathrm{m}42.44\mathrm{s}, +27^\circ56\mathrm{'}45.53\mathrm{''})$, and the other is referred to as the ``analysis centre'' located at $(\alpha, \delta) = (13\mathrm{h}00\mathrm{m}18.043\mathrm{s}, +28^\circ06\mathrm{'}46.07\mathrm{''})$.
The analysis centre was defined in such a way that a circle of radius 40 arcmin centred on this point traces most closely the western edge of the 352 MHz WSRT radio contours (see yellow dotted curve in Figure \ref{fig:w1234}).

\subsection{Spectral Analysis}\label{specan}

To derive the temperature profile, we extracted and modelled spectra from six annuli with radii 26--30, 30--35, 35--40, 40--45, 45--52, 52--60 arcmin from the cluster centre,
and 30--35, 35--40, 40--45, 45--50, 50--55, 55--62 arcmin from the ``analysis centre''. Visually identified point sources or compact substructures marked by white circles in Figure \ref{fig:w1234} were excluded from the spectral extraction regions.

The spectral modelling was performed with the XSPEC spectral fitting package (version 12.8.2, \cite{b37}), employing the modified C-statistic estimator. 
Unless otherwise noted, statistical errors are reported at the $\Delta C=1$ confidence level.
The spectra were binned to a minimum of one count per channel and fitted in the 0.7--7.0 keV band. As the spectral model, we employed an absorbed thermal plasma in collisional ionization equilibrium using the {\it apec} code (\cite{b22}), in addition to the X-ray background model described in Section \ref{sect:xrb}. 

The free parameters in the spectral model are the normalization and temperature of the {\it apec} component for each annulus. The {\it apec} metal abundance was treated as a free parameter but coupled to have the same value for all annuli. We obtain a best-fit value of $Z = 0.20\pm 0.07$ Solar in the units of \cite{b24}. Within the $1\sigma$ statistical confidence interval, this value is consistent with that reported by \cite{wer13} for the outskirts of the Perseus Cluster.

\subsection{Temperature profile}

\begin{table*}
\centering
\caption{Temperatures, normalizations, and column densities of the Galactic absorption for each annulus.
The unit of the spectrum normalization is defined as $\int n_\mathrm{e}n_\mathrm{H} \mathrm{d} V \times( 10^{-14}/(4\pi[D_A(1+z)]^2)) \mathrm{\ cm}^{-5}$ per $20^2 \pi$ arcmin$^2$. }
\begin{tabular}{cccc} \\\hline
Radius (arcmin) & $kT$ (keV) & $\mathit{Norm}$ $(\times 10^{-2})$ & $N_\mathrm{H}$ ($10^{19}$ cm$^{-2}$) \\ \hline
\multicolumn{4}{c}{centering on the cluster centre} \\ \hline
26 - 30 & $9.19^{+0.71}_{-0.71}$ & $4.40 \pm 0.07$ & $8.73$ \\
30 - 35 & $8.67^{+0.76}_{-0.54}$ & $2.47 \pm 0.04$ & $8.73$\\
35 - 40 & $10.67^{+2.26}_{-1.27}$ & $1.20 \pm 0.03$ & $8.61$\\
40 - 45 & $4.32^{+0.76}_{-0.62}$ & $0.69 \pm 0.04$ & $8.48$\\
45 - 52 & $5.39^{+0.91}_{-0.73}$ & $0.67 \pm 0.03$ & $8.48$\\
52 - 60 & $3.36^{+1.32}_{-0.81}$ & $0.26^{+0.03}_{-0.02}$ & $8.48$\\\hline
\multicolumn{4}{c}{centering on the analysis centre} \\ \hline
30 - 35 & $8.59^{+0.80}_{-0.46}$ & $4.61 \pm 0.07$ & $8.73$ \\
35 - 40 & $9.40^{+0.71}_{-0.71}$ & $2.81 \pm 0.04$ & $8.73$\\
40 - 45 & $8.94^{+1.11}_{-1.03}$ & $1.28 \pm 0.03$ & $8.61$\\
45 - 50 & $5.33^{+1.14}_{-0.91}$ & $0.60 \pm 0.03$ & $8.48$\\
50 - 55 & $5.27^{+0.91}_{-0.65}$ & $0.58 \pm 0.02$ & $8.48$\\
55 - 62 & $4.65^{+0.78}_{-0.65}$ & $0.42 \pm 0.02$ & $8.48$\\\hline
\end{tabular}
\label{tab:ktnorm}
\end{table*}

\begin{figure}
\begin{center}
\vspace{0.7cm}
\includegraphics[angle=270,width=3.2in]{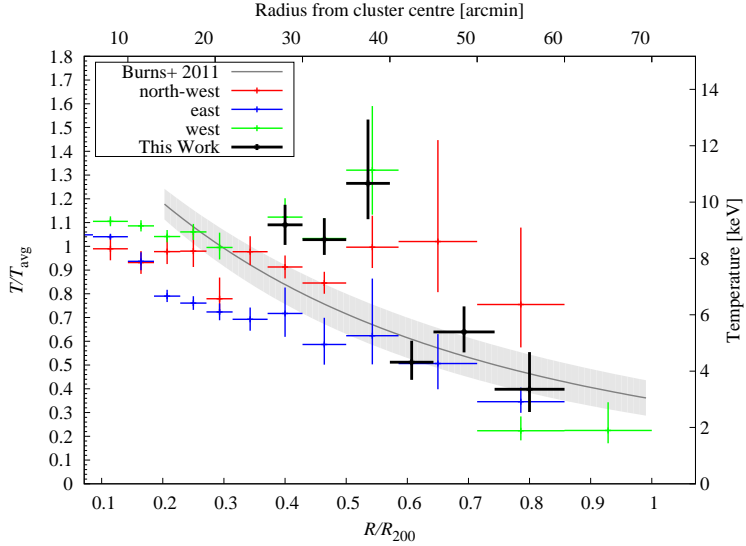}
\vspace{0.7cm}
\end{center}
\caption{Suzaku temperature profile for the Coma Cluster. Black points were obtained in this work. The other measurements, shown with dashed lines, are reproduced from \cite{b2}. Green, red and blue represent the west, north west and east directions, respectively. The gray curve indicates the universal profile expected from numerical simulations of relaxed clusters (\cite{b14}) scaled to the average temperature of this system. ($R_{200}  = 70$ arcmin.)}
\label{kt_a}
\end{figure}

\begin{figure}
\begin{center}
\includegraphics[angle=270,width=3in]{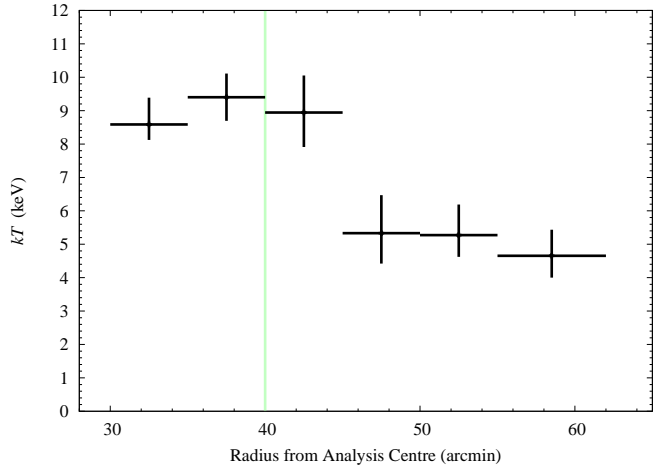}
\vspace{0.8cm}
\end{center}
\caption{Temperature profile centred on the analysis centre. A discontinuity can be confirmed around 45 arcmin. The green line indicates the location of the radio halo edge.}
\label{kt_b}
\end{figure}

The values for the best-fit parameters obtained from the spectral fitting are summarised in Table \ref{tab:ktnorm}. Figures \ref{kt_a} and \ref{kt_b} show the radial profiles of the temperature obtained from this analysis for the two assumed sets of annuli.

The temperature profile obtained from annuli centred on the cluster centre shows a marked temperature jump between the annuli spanning $35-40$ arcmin and $40-45$ arcmin (Figure \ref{kt_a}). This indicates the presence of a shock front. 

From the Rankine-Hugoniot shock jump condition (see \cite{b15,b7}), the Mach number can be estimated from the temperature discontinuity using the following equations:
\begin{equation}
\mathcal{M} = \left( \frac{2\rho}{\gamma+1-\rho(\gamma-1)} \right)^{1/2} , 
\end{equation}
\begin{equation}
\rho^{-1} = \left(\frac{1}{4} \zeta^2 ( t-1)^2 + t \right)^{1/2} - \frac{1}{2} \zeta (t-1),
\end{equation}
where $\gamma=5/3$ is the adiabatic index, $\rho \equiv \rho_1/\rho_0 $ is the electron density jump, $t \equiv T_1/T_0$ is the gas temperature jump, 
and $\zeta \equiv (\gamma +1)/(\gamma-1)$. 
The indices 0 and 1 indicate the values of the respective quantities before and after the shock passage, respectively.
Based on the observed temperature jump and taking into account the propagation of errors, we estimated the Mach number as $\mathcal{M} = 2.3 \pm 0.4$.

However, the Mach number determined in this way is only valid under the assumption that the entirety of the temperature difference between the annuli at $35-40$ arcmin and $40-45$ arcmin is due to shock heating or -- otherwise put -- that no temperature gradient between these regions was present before the shock passage. In reality, the temperature is expected to decline with radius in the cluster outskirts, even in the absence of a shock.
From $N$-body simulations, \cite{b14} predict that, for relaxed clusters, the temperature profile as a function of radius can be described by the following equation:
\begin{equation}
\frac{T}{T_\mathrm{avg}} = A\left[1+B\left(\frac{r}{r_{200}}\right)\right]^{-\beta},
\end{equation}
where $T_\mathrm{avg}$ is the average X-ray weighted temperature between 0.2 and 1.0 $r_{200}$. 
\cite{b14} report the best-fit values of the other parameters as $A = 1.74 \pm 0.03$, $B = 0.64 \pm 0.10$ and $\beta = 3.2 \pm 0.10$.
Using the measurements along all azimuths of the Coma Cluster covered by Suzaku (this work and \cite{b2}), we obtain $T_\mathrm{avg} = 8.43\pm 0.05$ keV, resulting in the predicted temperature profile shown as a grey band in Figure \ref{kt_a}. The measured temperature jump at the 40 arcmin shock is clearly higher than the temperature gradient expected for a relaxed cluster. The pre-shock temperature in the $35-40$ arcmin annulus predicted from the universal temperature profile of \cite{b14} is $T_\mathrm{univ.} = 5.71 \pm 1.0$ keV, which would imply a Mach number $\mathcal{M}=1.8 \pm 0.4$ -- smaller than, but roughly consistent with the value derived without accounting for the underlying temperature gradient in the absence of a shock.

We also evaluated the temperature profile from annuli centred on the ``analysis centre'' (see Figure \ref{kt_b}). In this case, the temperature jump appears between the annuli at 40--45 and 45--50 arcmin, and the corresponding Mach number is estimated as $\mathcal{M} = 1.7 \pm 0.3$. Note that, in this case, due to the special geometry, we cannot apply the same correction for the underlying temperature gradient before the passage of the shock as discussed in the previous paragraph and our value is therefore an upper limit for the measured Mach number.

\subsection{Systematic uncertainties on the temperature estimation}

In order to account for the systematic uncertainties associated with the X-ray background model, we divided the area beyond the cluster's $r_{200}$ from which the average model parameters were estimated by \cite{b2} into 16 different spectral extraction regions, and fitted each region separately with the normalizations of the CXB power-law and GH thermal emission allowed to vary. All other parameters of the X-ray background model were kept fixed to the values described in Section \ref{sect:xrb}. Each such background region roughly corresponded to a half of the Suzaku XIS field of view.

\begin{figure}
\begin{center}
\includegraphics[angle=270,width=3in]{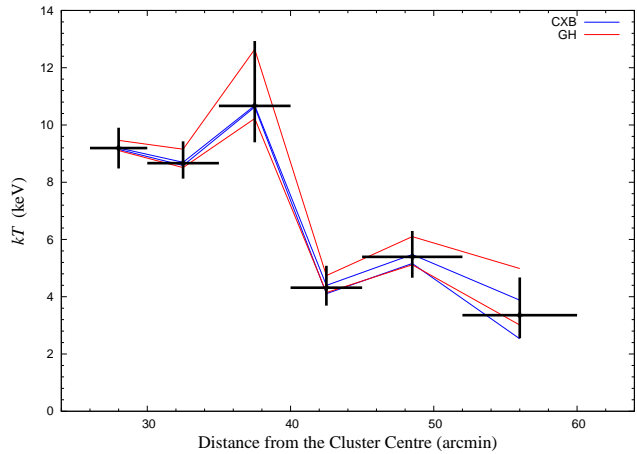}
\vspace{0.7cm}
\end{center}
\caption{The expected effects of spatial variations of the CXB and GH normalisations on the measured radial temperature profile. The blue and red lines indicate temperature changes by varying the CXB and the GH normalizations within their $1\sigma$ confidence intervals, respectively.
}
\label{cxb_gh}
\end{figure}

We then calculated the mean and standard deviations of the CXB and GH normalisations from these 16 different regions and refitted the radial temperature profile by setting each parameter in turn to its mean$-1 \sigma$ and mean$+1 \sigma$ values. In practice this amounts to a $\pm$9\% variation of the CXB and $\pm$30\% variation of the GH fluxes. 
Fig. \ref{cxb_gh} shows the results. We conclude that the effect of systematic uncertainties is small. This is not unsurprising, since the shock is located at radii of only about 40 arcmin, less than two-thirds of the Coma Cluster's $r_{200}$.


\subsection{Surface brightness profile}

The presence of a shock feature in the ICM should cause not only a temperature jump, but also a discontinuity in the gas density and therefore the X-ray surface brightness. To search for such a density jump associated with the shock, we extracted surface brightness profiles from the XIS mosaic shown in Figure \ref{fig:w1234}, {using the energy band between 0.7 and 7.0 keV.}

We then used the PROFFIT code v1.1 (\cite{b20}) to fit the surface brightness profiles with a projected broken power-law model, assuming spherical symmetry with respect to the centre of the annuli used to obtain the radial profile:
\begin{equation}
n_1 = C n_{\rm norm} \left(\frac{r}{r_\mathrm{break}}\right)^{-\alpha_1}, \ \ r < r_\mathrm{break} 
\end{equation}
\begin{equation}
n_0 =  n_{\rm norm} \left(\frac{r}{r_\mathrm{break}}\right)^{-\alpha_0}, \ \ r > r_\mathrm{break}.
\end{equation}
In the above equations, $n$ is the electron number density, $C$ is the density compression factor, $r$ is the distance from the centre of the sector, $r_\mathrm{break}$ is the break radius and $\alpha$ is the power-law index. Subscripts 1 and 0 indicate the post-shock and pre-shock properties, respectively.

To obtain the fitting results, we adopted the Cash statistics, which is appropriate for the low-count regime. We include the sky background as an additional constant surface brightness component in our model. The value of this constant was estimated from the same Suzaku pointings that were used to determine the X-ray background model used for spectral analysis (these background fields are located beyond the 70 arcmin, the cluster's estimated $r_{200}$). 
We include the effects of Suzaku's point spread function (PSF) by convolving the surface brightness model with a Gaussian with a full width at half maximum of 1.9 arcmin.

\begin{figure*}
\begin{center}
\includegraphics[angle=270,width=3.0in]{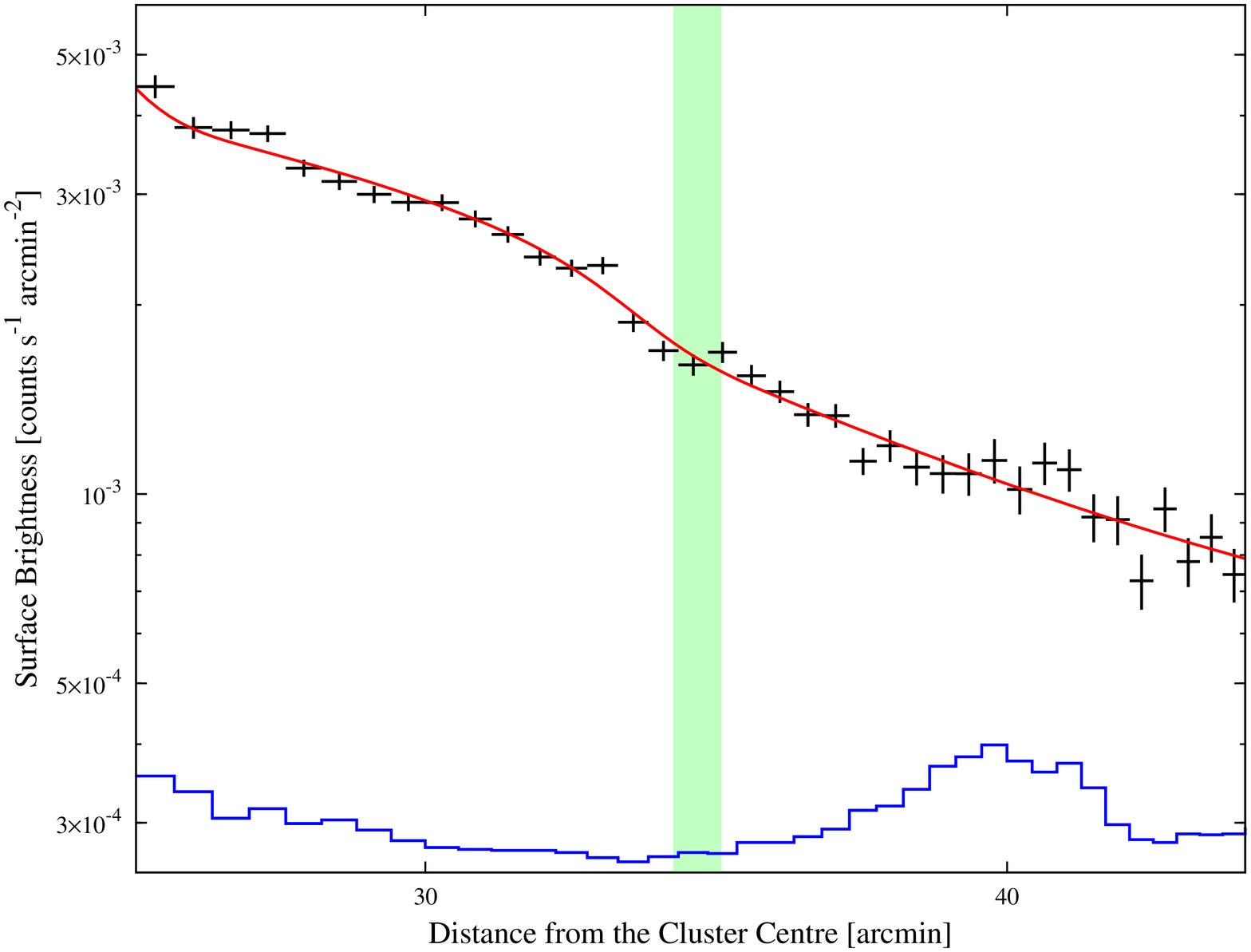} 
\hspace{30pt}
\includegraphics[angle=270,width=3.0in]{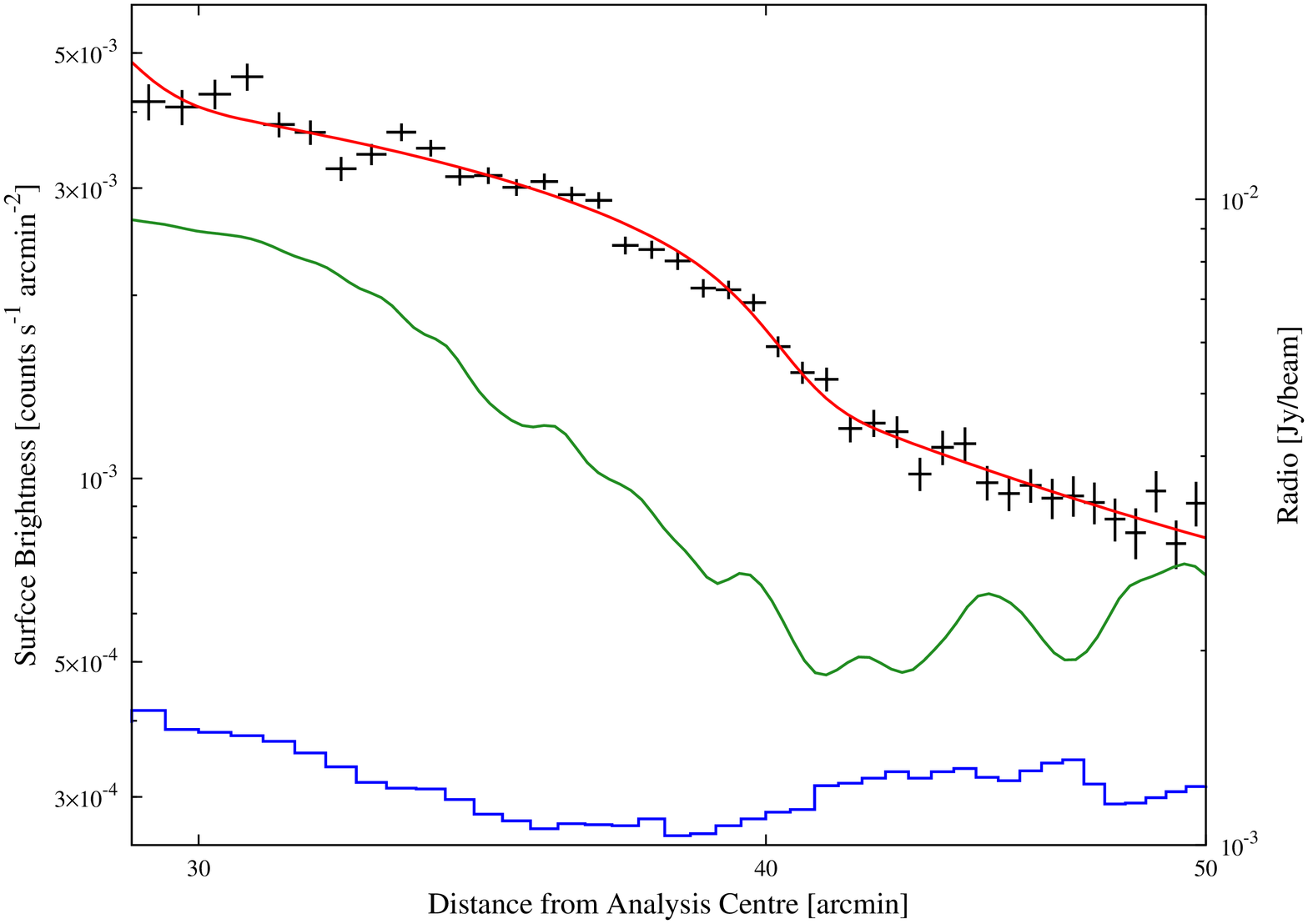}
\vspace{0.7cm}
\end{center}
\caption{Surface brightness profiles of the Coma Cluster and the best-fit projected broken power-law models centred on the cluster centre (left) and the ``analysis centre'' chosen to match the curvature of the outer 352 MHz WSRT radio contours (right). In both panels, the red lines indicate the broken power-law model and the blue lines indicate the instrumental background, NXB. The green vertical shaded region in the left panel indicates the radial range of the radio halo edge. In the right panel, a dark-green line indicates the surface brightness profile obtained from the 352 MHz map.}
\label{sbprof_a}
\end{figure*}

We first fitted the radial surface brightness profiles centred on the cluster centre, restricting our fit to the radial range between 26 and 45 arcmin. The result is shown in Figure \ref{sbprof_a}.
The best fit parameters are 
$\alpha_0 = 0.76^{+0.05}_{-0.06}$, 
$\alpha_1 = 2.59 \pm 0.13$, 
$r_\mathrm{break} = 33.4^{+0.26}_{-0.30}$ arcmin and 
$C = 1.31 \pm 0.06$.
Accordingly, the corresponding Mach number from the Rankine-Hugoniot condition for the density jump is $\mathcal{M} = 1.21 \pm 0.04$.

For fitting the surface brightness profiles centred on the ``analysis centre'', we used the radial range between 29 and 50 arcmin. The density discontinuity is stronger and more easily seen, and the computed best fit parameters are 
$\alpha_0 = 0.41^{+0.08}_{-0.09}$, 
$\alpha_1 =2.59^{+0.23}_{-0.26}$, 
$r_\mathrm{break} = 40.5 ^{+0.22}_{-0.21}$ arcmin and 
$C = 1.65^{+0.12}_{-0.09}$, resulting in the Mach number $\mathcal{M} = 1.45 \pm 0.08$.
As can be seen in the right-hand panel of Figure \ref{sbprof_a}, the location of the edge in the X-ray surface brightness profile computed with respect to the ``analysis centre'' corresponds well to the location of the radio halo's edge.

We have also searched for spatial variations of the Mach number by calculating surface brightness profiles centred on the ``analysis centre'' in two separate wedges with equal opening angles. We find no evidence for significant differences in the shock strength, with $r_\mathrm{break} = 40.8\pm0.4$ arcmin and $\mathcal{M} = 1.47 \pm 0.15$ for the northern half and $r_\mathrm{break} = 40.4\pm0.5$ arcmin and $\mathcal{M} = 1.32 \pm 0.10$ for the southern half of the shock covered by Suzaku.


\section{Discussion and Conclusions}

We have analysed data from four Suzaku pointings located along the western side of the Coma Cluster. Both the temperature and surface brightness distributions reveal the presence of a shock front likely associated with the edge of the Coma radio halo. 
In this section, we investigate the strength and location of the shock, in comparison with previous results estimated from X-ray, SZ, and radio observations, and discuss particle acceleration at X-ray shocks in relation to the origin of the radio emission.

\subsection{Shock strength, location, and geometry}

Previous X-ray studies have already revealed the presence of temperature structure in the Coma Cluster.
ASCA observations (\cite{b26,b27}) showed a temperature asymmetry, with the western side of the cluster being hotter than the eastern side. 
This trend was later confirmed by XMM-Newton and Suzaku (e.g. \cite{b28,b2}).
This temperature structure is thought to be associated with a past merger (\cite{b26,b28}). 

Using SZ observations, \cite{b1} report the presence of two shock fronts located about 30 arcmin towards the west and south-east from the cluster centre, with Mach numbers $\mathcal{M}_W=2.03^{+0.09}_{-0.04}$ and $\mathcal{M}_{SE}=2.05^{+0.25}_{-0.02}$ derived from the Rankine-Hugoniot pressure jump conditions. Both of the shocks reported by \cite{b1} align well with the sharp gradients in the 352 MHz WSRT radio brightness distribution, and are associated with X-ray surface brightness edges and temperature jumps detected in XMM-Newton data. Thus, shock heating is likely at least partly responsible for the elevated gas temperatures on the western side of the cluster reported by previous studies.

In this work, we also report the presence of a shock front along the western side of the Coma Cluster, although located further north than the western shock reported by \cite{b1}; the area corresponding to the shock seen by Planck is not covered by Suzaku observations. For different choices of extraction regions and different assumptions about the underlying temperature profile in the absence of shock heating, the temperature profiles measured with Suzaku imply Mach numbers ranging from $\mathcal{M} = 1.7 \pm 0.3$ to $\mathcal{M} = 2.3 \pm 0.4$, broadly consistent with the Mach number derived from SZ observations for what can be assumed to be a different azimuth along the same shock front that traces the entire western edge of the giant radio halo.

Shock signatures are seen not only in the temperature but also in the X-ray surface brightness distribution along this azimuth. When extracting radial profiles in annuli roughly parallel to the steep radio edge, we obtain consistent Mach numbers from the inferred temperature and surface brightness jumps, with $\mathcal{M} = 1.7 \pm 0.3$ and $\mathcal{M} = 1.45 \pm 0.08$, respectively. However, when choosing annuli centred on the peak X-ray emission of the Coma Cluster, the shock strengths derived from the temperature and surface brightness become inconsistent, with the Mach number inferred from the temperature jump, $\mathcal{M} = 2.3 \pm 0.4$, being larger than that determined from the density jump, $\mathcal{M} = 1.21 \pm 0.04$. Even when accounting for the expected underlying radial temperature gradient between the pre- and post-shock regions in a typical relaxed cluster, the Mach number inferred from the temperature jump remains high, at $\mathcal{M} = 1.8 \pm 0.4$. 

It is generally expected that a mis-alignment of the shock front compared to the orientation of the spectral extraction regions would lead to an underestimation of both the true temperature as well as density jumps. Therefore, the observed decrease of the Mach number estimated from the surface brightness profile, from $\mathcal{M}= 1.45 \pm 0.08$ to $1.21\pm0.04$ for a less optimised geometry, is easily understood; however, assuming the shock has a spherically symmetrical shape, it is difficult to explain why the shock strength inferred from the temperature jump does not show a commensurate decrement. 


Perhaps even more surprisingly, the locations of the discontinuities determined from the X-ray temperature and surface brightness differ. When considering the profiles centred around the ``analysis centre'', we obtain $r_\mathrm{break} = 40.5 ^{+0.22}_{-0.21}$ arcmin from the surface brightness profile fitting; this is in good agreement with the location of the radio halo edge (the ``analysis centre'' was chosen so that a circle of radius 40 arcmin centred on this point traces most closely the western 352 MHz WSRT radio contours). On the other hand, the temperature jump is seen at a radius of 45 arcmin; a quick inspection of Figure \ref{kt_b} reveals that, for a shock located at a radius of 40.5 arcmin, we expect to detect a temperature jump between the second and third annuli (35--40 and 40--45 arcmin), rather than between the third and fourth annuli as currently seen. A similar conclusion is reached for profiles centred around the cluster centre. In this case, the surface brightness jump is seen at $33.4^{+0.26}_{-0.30}$ arcmin, in good agreement with the location of the radio edge contours spanning 34--35 arcmin with respect to this choice of centre, while the temperature jump is seen at 40 arcmin. 
 
Thus, the locations of the temperature and density jumps appear to differ by 4.5--6.6 arcmin (125--185~kpc), with the temperature jump located further out while the surface brightness jump agrees well with the position of the edge of the radio halo. This difference is significantly larger than the Suzaku PSF, and is seen regardless of the choice of the centre used to extract the profiles. If the locations of the temperature and density jumps are better aligned for the southwestern half of the region characterised by a steep radio gradient, this would explain why the northwestern shock has not been previously identified either by \cite{b1} (the pressure jump is smeared because the product of temperature and density now has discontinuities at multiple radii), or by \cite{b11} who showed that XMM-Newton spectra from smaller extraction regions on either side of the radio edge only suggest a significant temperature jump towards the southwest but not towards the northwest.

A possible explanation for the different shock radii inferred from the temperature and surface brightness profiles is that the shocked gas that covers the largest line of sight depth (and has contributed to accelerating the most radio-emitting particles) has been heated by a weaker shock and lies closer to the centre, while the temperature jump is driven by gas affected by a stronger shock at slightly larger radii, but where the shock has a smaller line of sight depth; while this might still steepen the power-law fitted to the surface brightness distribution, the line-of-sight effect may conceal an actual discontinuity corresponding to the location of the temperature jump. We note that numerical simulations do indeed suggest both complex shapes as well as complex distributions of the Mach number along the shock fronts believed to be responsible for accelerating particles to the relativistic energies required to produce radio emission in radio relics (\cite{skillman2013}), so it is plausible that this may also apply for the Coma radio halo.
An alternative, perhaps more exotic scenario, is that in which a fraction of the particles accelerated by the shock may diffuse ahead of the shock and cool out of the radio band by mixing with and heating the thermal ICM.

\subsection{Particle acceleration at X-ray shocks and the origin of the radio emission}

The two main hypotheses typically invoked in the past in order to explain the synchrotron emission in radio haloes are typically (re)acceleration of electrons up to relativistic energies due to the turbulence induced by cluster mergers (e.g. \cite{b35}), or collisions between thermal ions and relativistic protons in the ICM (\cite{b36}). Concomitantly, in a growing body of publications, the edges of radio haloes are found to be coincident with shock fronts (e.g. \cite{b8}, \cite{b9}; \cite{b30}, \cite{b1}, and this work). This suggests that shocks are an additional process contributing to the acceleration of particles in radio haloes. The clusters where shocks are seen to be associated with the edges of radio haloes therefore present a unique angle to study a superposition of particle acceleration mechanisms acting simultaneously. The Coma Cluster is the nearest, brightest, and to date one of the best studied among this type of systems. 

One of the main theoretical challenges to be addressed in this case is the exact mechanism through which very weak shocks ($\mathcal{M} \sim 2$ in Coma and typically $<3$ in other clusters cited above) are in fact able to boost the radio surface brightness. For such low Mach numbers, the diffusive shock acceleration (DSA) mechanism alone is thought to be inefficient at accelerating particles. The preferred interpretation is that, instead of directly accelerating the thermal electrons, the shocks re-accelerate a population of already mildly relativistic electrons present in the cluster (e.g. \cite{b9}, \cite{b39}, \cite{b40}). 

The morphology of the Coma radio halo supports this interpretation; the observed steepening, or ``edge'', in the radio surface brightness profiles reported by \cite{b11}, is sharpest along the western side of the cluster, and is not apparent along most other directions. Suzaku demonstrates for the first time that a shock front extends not only along the southern but also along the northern half of this radio brightness edge. Therefore, we can infer that the difference in radio brightness between the western side of the cluster and other azimuths is entirely attributed to reacceleration due to the shock passage.

Recent particle-in-cell simulations indicate that a different mechanism termed shock drift acceleration may result in a more efficient acceleration at low-Mach number shocks compared to standard DSA (e.g. \cite{b41,b42,b43}), further boosting the radio surface brightness at the western Coma shock.

\section*{Acknowledgments}
The authors thank the Suzaku operation team and Guest Observer Facility, supported by JAXA and NASA. We are grateful to L. Rudnick for providing the 352 MHz WSRT map and to G.A. Ogrean, J. Zrake, and G. M. Madejski for helpful discussions about this manuscript. YI is financially supported by a Grant-in-Aid for Japan Society for the Promotion of Science (JSPS) Fellows. 

\appendix
\section{XIS spectra}

\begin{figure*}
\includegraphics[width=2.2in]{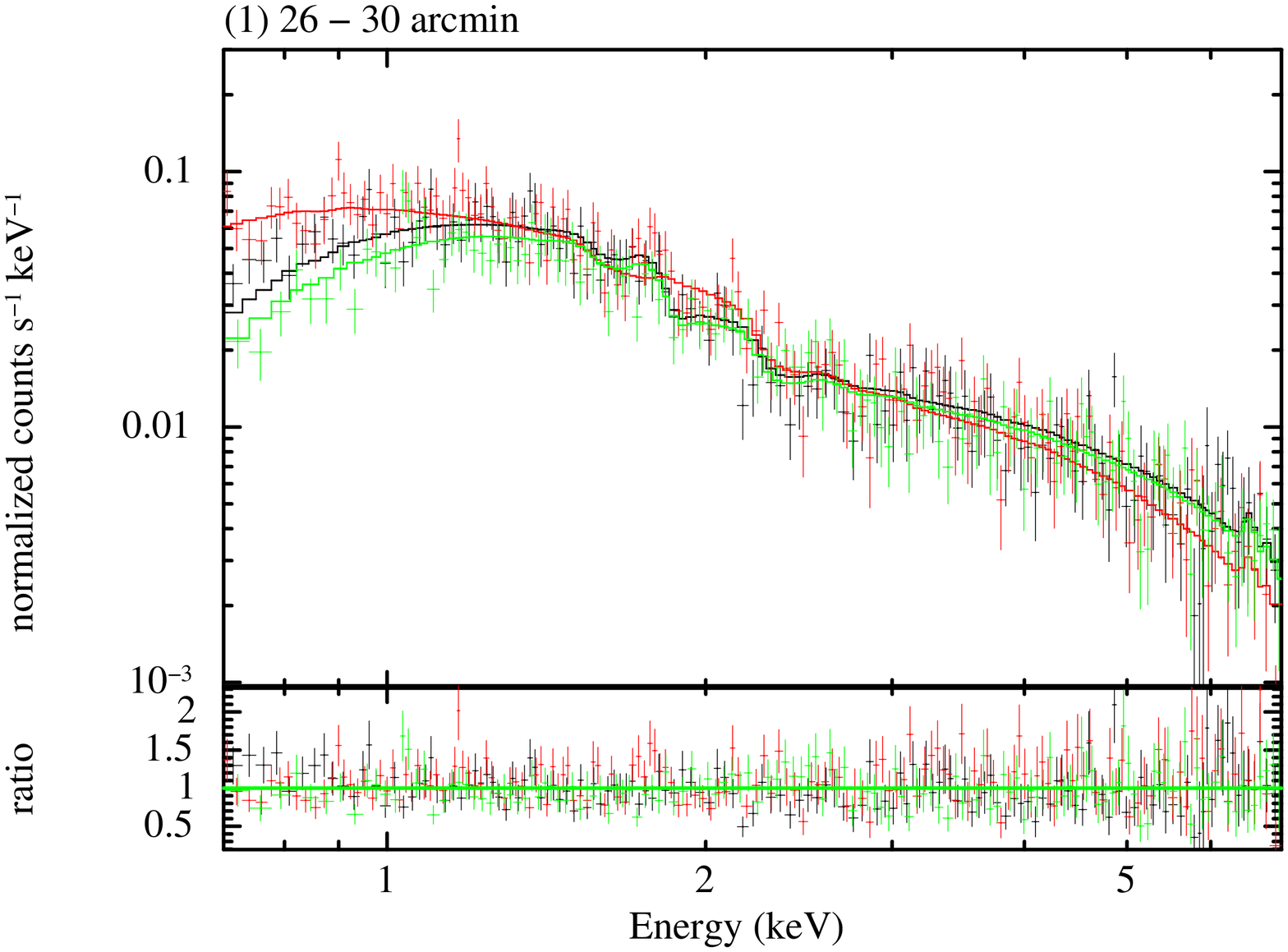}
\includegraphics[width=2.2in]{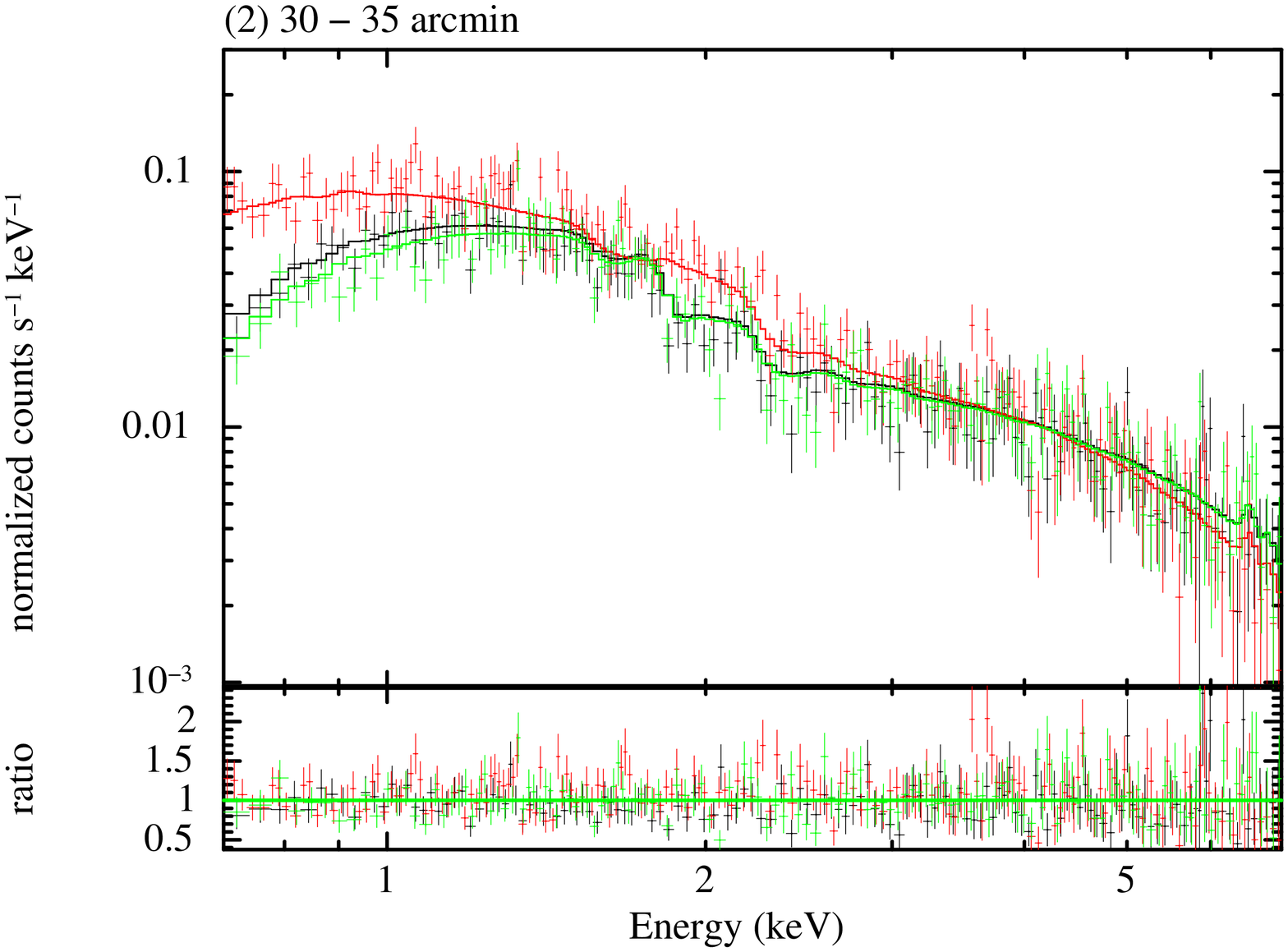}
\includegraphics[width=2.2in]{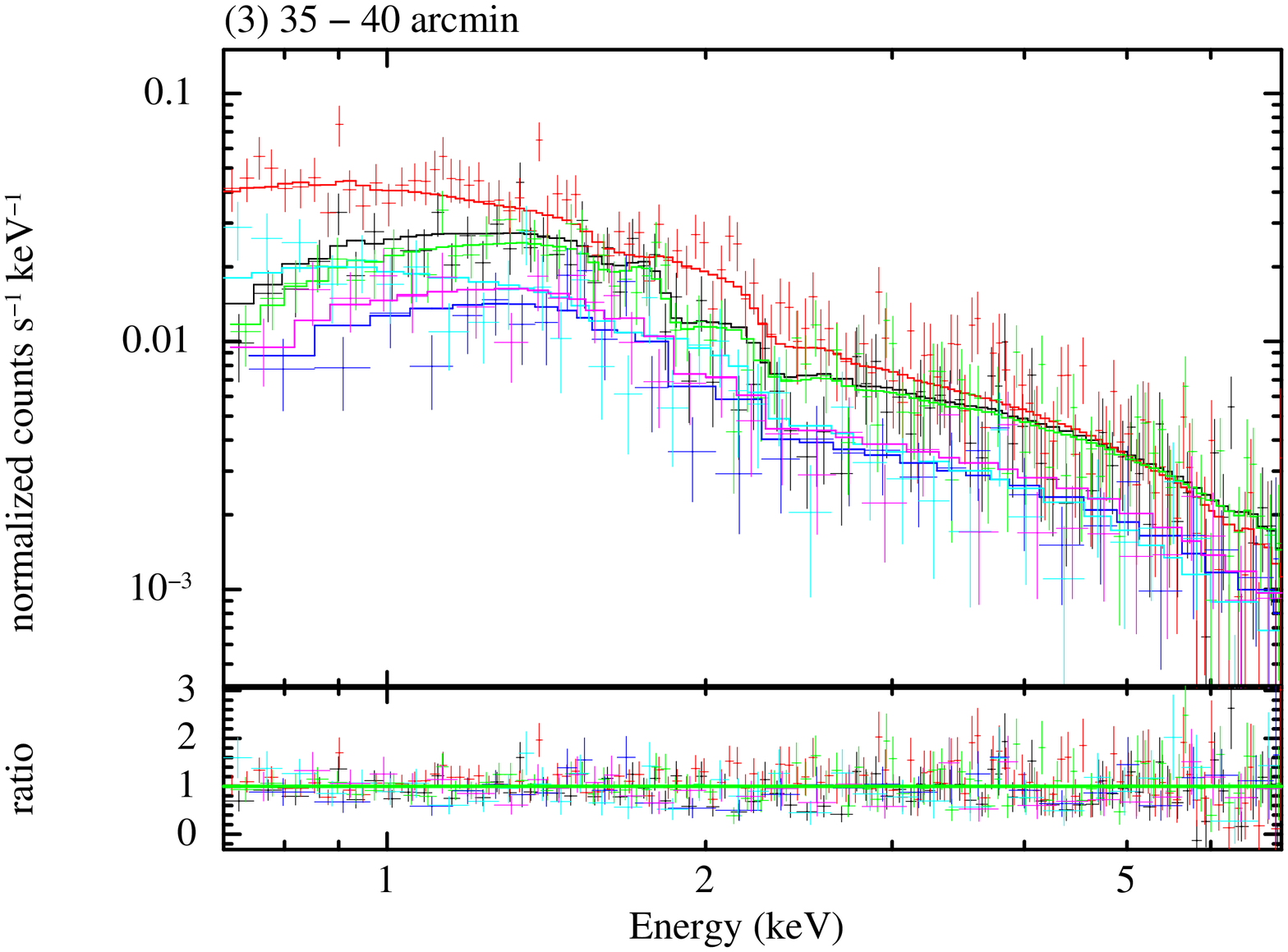}
\includegraphics[width=2.2in]{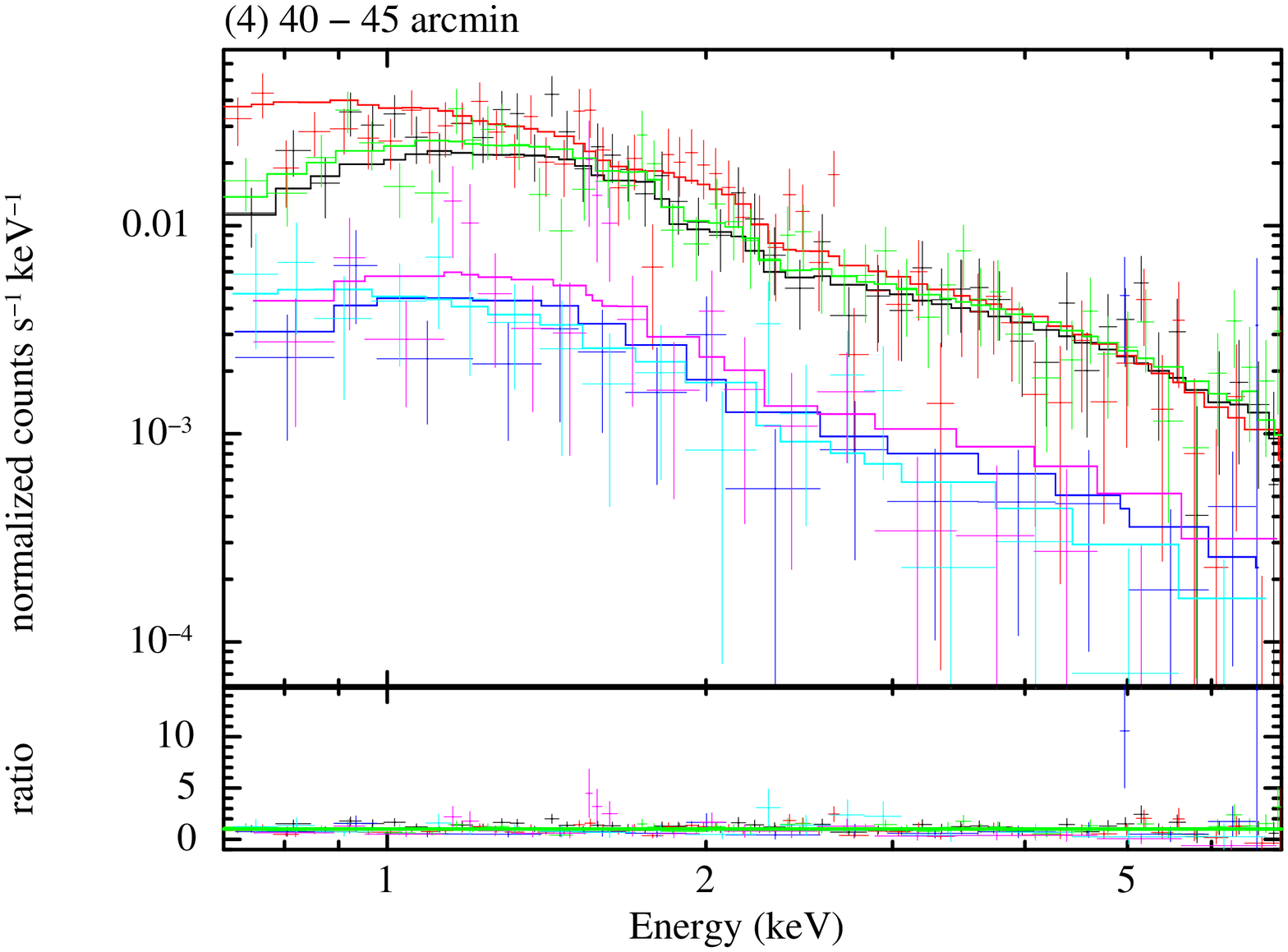}
\includegraphics[width=2.2in]{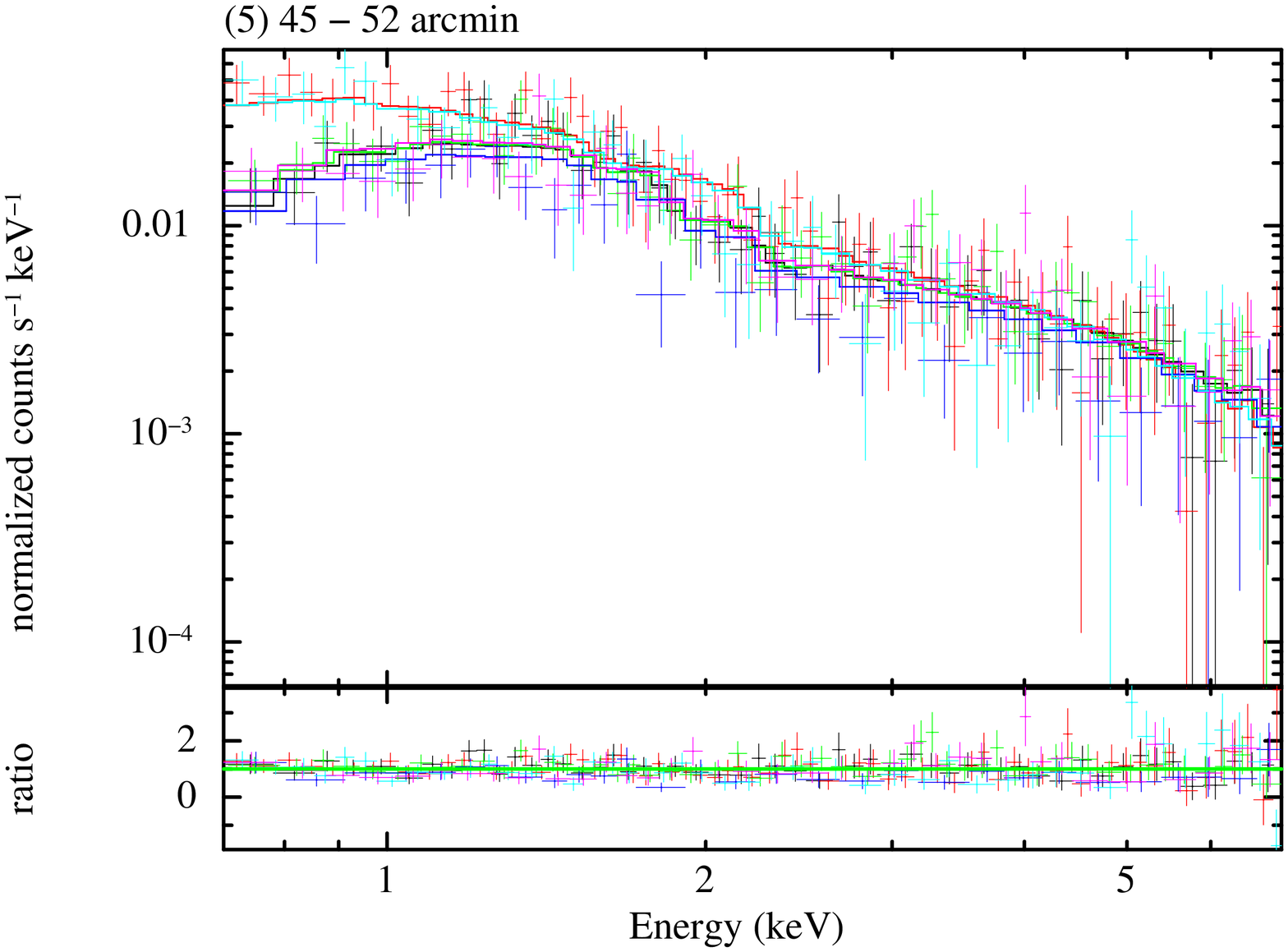}
\includegraphics[width=2.2in]{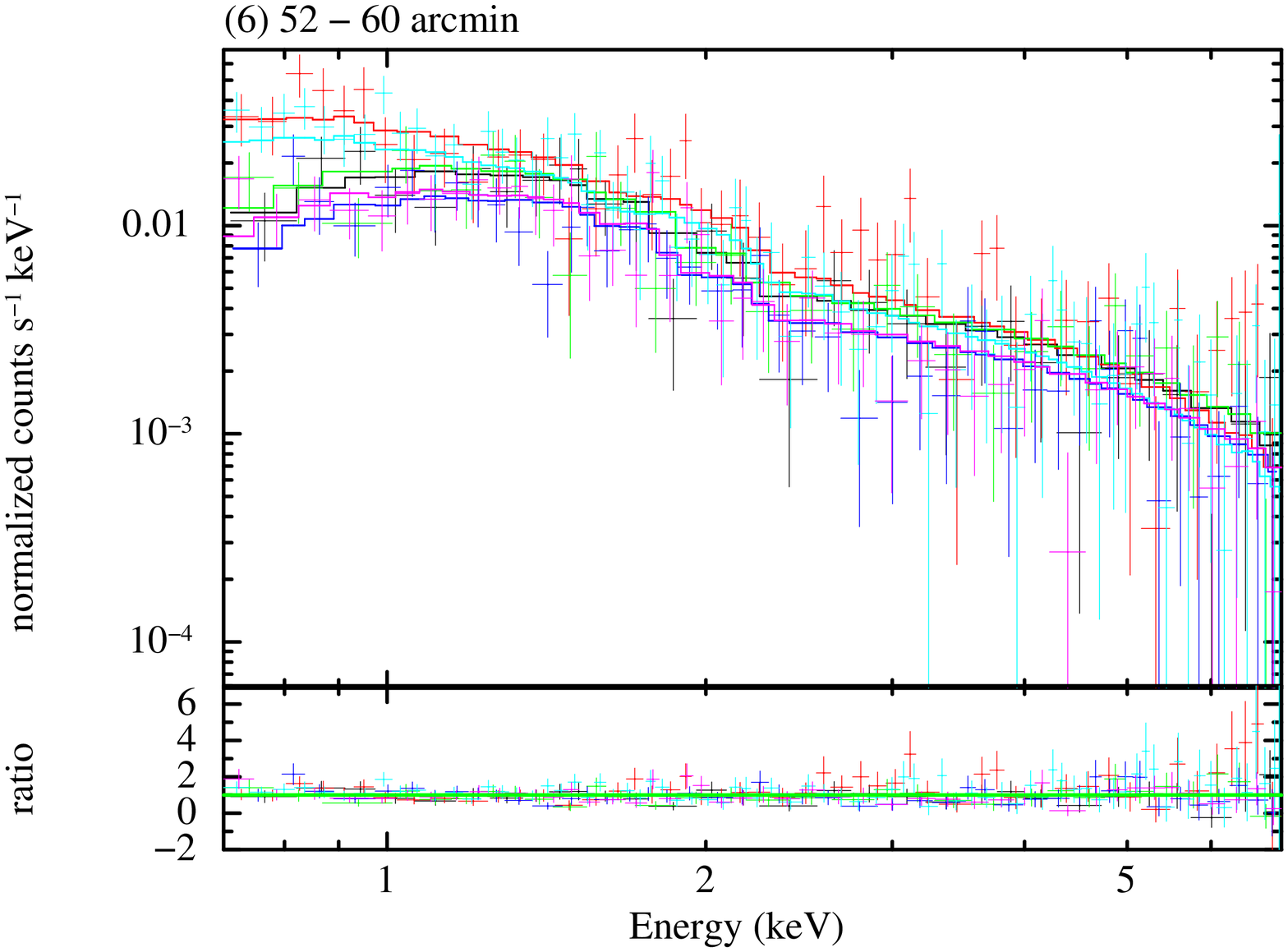}
\caption{Observed XIS spectra for each annulus centered on the cluster center. (1) 26-30 arcmin (2) 30-35 arcmin (3) 35-40 arcmin (4) 40-45 arcmin (5) 45-52 arcmin (6) 52-60 arcmin. We model the spectra with a thermal component describing the ICM of the Coma Cluster (Section 4.2) in addition to a model describing the X-ray backgrounds (Section 3.2). The solid lines indicate the best fit model. We show separately the data from all three detectors and different overlapping pointings within each annulus: XIS0 (black and blue), XIS1 (red and cyan) and XIS3 (green and magenta). }
\label{sb}
\end{figure*}

\end{document}